\documentclass[aps,pra,preprint,showpacs]{revtex4}

\usepackage{graphicx}
\usepackage{amsmath}
\usepackage{bm}

\begin{document}

\title{Muonic atoms with extreme nuclear charge}

\author{V. V. Flambaum}
 \affiliation{School of Physics, University of New South Wales, Sydney, 2052 Australia}
\author{W. R. Johnson}
\email{johnson@nd.edu} \affiliation{Department of Physics,  225
Nieuwland Science Hall, University of Notre Dame, Notre Dame, IN
46556}

\date{\today}
\begin{abstract}
Bound muons (also pions, kaons, etc) increase the fission barrier
and produce some stabilizing effects for highly charged nuclei. If
the binding energy of the muon exceeds $mc^2$, it becomes stable.
The $1s$ state of  a muon inside an exotic nucleus with atomic
number $A= 5Z/2$ and such large charge $Z$  that the $1s$ energy $E$
is in the range $0\le E<-mc^2$ is considered.
\end{abstract}
 \maketitle

    As well known, all atoms with nuclear charge $Z> 100$ have very short lifetimes
and atoms with $Z>118$ have not as yet been found.  However, the
behavior of orbitals  in the field of a highly charged ``nucleus''
remains an interesting theoretical problem. There are various
motivations for the study of such systems. Firstly, in collisions of
heavy nuclei  (e.g. uranium with charge $Z_i=92$)
 a compound nucleus with the charge $Z=2 Z_i$  is formed for a short time.
Electron-positron pair creation due to ``vacuum breakdown'' is expected in this
situation, provided the total compound nucleus charge $Z$ exceeds a critical
charge $Z_c \approx  170$ (see e.g. review \cite{ZeldovichPopov}
 and the book \cite{Greiner}).
An experimental investigation of this phenomenon has been done in GSI, with
new experiments in preparation \cite{GSI}.
Then, there are speculations about exotic forms of matter such as
``strange matter'' where a nucleus contains strange quarks in
addition to ``normal''  $u$ and  $d$ quarks \cite{Witten}; there
may be highly charged small primordial black holes \cite{Hawking}
and the associated black hole ``atoms''
\cite{Berengut}.

  In this note, we would like to consider the purely theoretical problem
of a charged particle in the superstrong Coulomb field. The electron
problem is actually very well studied (see e.g. \cite{ZeldovichPopov}).
 However, the Coulomb problem for a heavier particle has some new
and interesting features.
Consider, for example, a muon. Normally, it decays into
electron and neutrinos. However, the bound muon is stable if the binding energy
exceeds $mc^2$ so the total relativistic energy $E<0$.
 In this case the bound muon has a significant stabilizing
effect on the ``nucleus'', it contributes $\sim mc^2$ $\approx  100$
MeV to the nuclear fission barrier.  In the case of bosonic particles
like pions or kaons the effect can be even more significant since
many particles may be placed in $1s$ state and the critical charge for spin
 $S=0$ and $S=1$ is much  smaller than for $S=1/2$ ($Z_c$=68 for a very
 small nucleus, i.e.\ a factor of two smaller than for Dirac particles).
 Moreover, the
 strong interaction may produce an additional attraction and reduce the
critical charge.

\begin{figure*}
\centerline{ \includegraphics[scale=0.8] {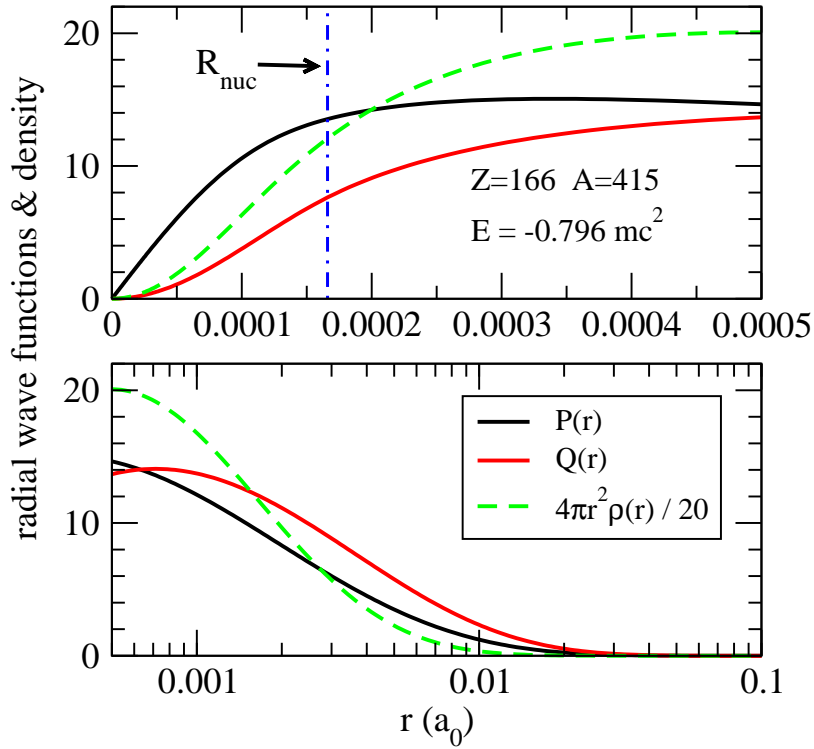} }
\vspace{1.5pc}

\caption{ Radial wave functions $P(r)$ and $Q(r)$ and the radial
charge density $4\pi r^2 \rho (r)$ of a $1s$ electron
($E= -0.796\, mc^2$) in the field of a nucleus of charge $Z=166$  are shown.
\label{fig1}}

\end{figure*}

 In our numerical calculations, we assume a standard nuclear charge density
with radius $R =1.15 A^{1/3}$ fm,  and atomic number $A=5Z/2$.
 Let us start our consideration from the conventional electron case.
 The electron density is localized mostly
outside the nucleus even as the nuclear charge approaches the
critical  value $Z_c \approx  170$;  thus,  there is no collapse
of the electron wave function inside the nucleus.  In Fig.~\ref{fig1}, we
plot the large and small components of $1s$ wave function
and the radial density  for charge $Z$=166,
 which is  close to the critical charge.  The electron energy in this case is
$E=-0.796\, mc^2$  (the energy for the critical charge is $E_c=-mc^2$).
Thus, the electron wave function for the finite nucleus  in the main
area is not very different from the point-nucleus case. Consequently,
 the critical charge
 for the  finite nucleus $Z_c \approx 170$ is relatively close to the
 case of a very small nucleus where $Z_c=138$.

Now consider the muon case. The non-relativistic muonic atom
 is $m_{\mu}/m_e=200$ times smaller than the usual
electronic atom. In the relativistic case $E <0$, the muon wave
function for the $1s$ state is
 localized inside the nucleus where the electrostatic potential
has the oscillator shape
\[
 V(r) = -\frac{Ze^2}{R}\left(\frac{3}{2}- \frac{r^2}{2R^2}\right) .
\]
The critical charge for this oscillator potential is very different
from the Coulomb potential. The energy reaches $E=0$ for
a nuclear charge $Z\approx 840$ and the critical value $E=-mc^2$ for
$Z_c \approx 2000$.

\begin{figure*}
\centerline{\includegraphics[scale=0.8]{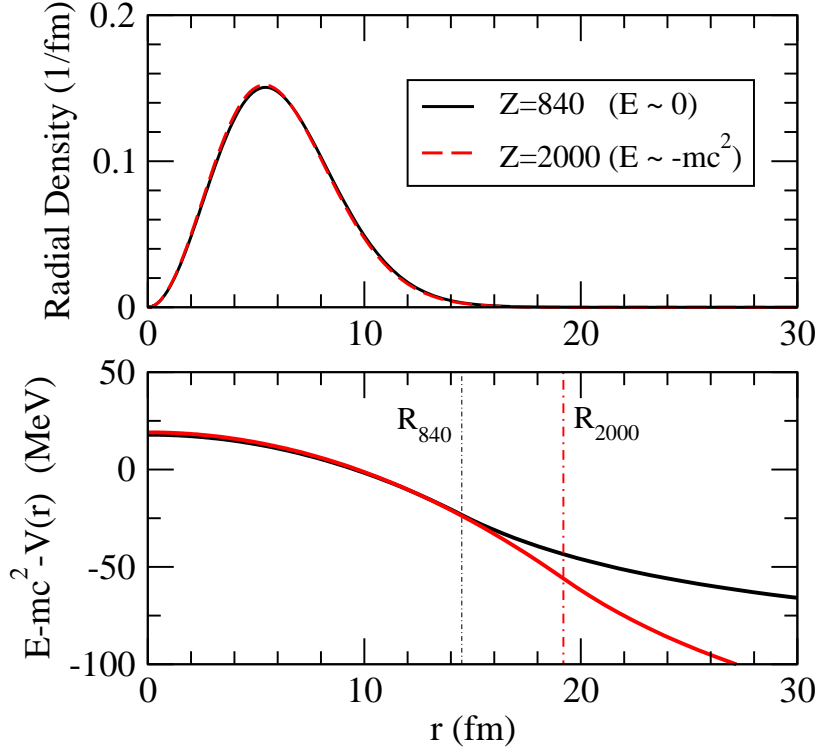}}
 \vspace{1.5pc}

\caption{Upper panel: The $1s$ radial density for a muon (energy
slightly above $-mc^2$)  in the field of a nucleus with charge
$Z=2000$ and radius $ R=19.2$~fm (red curve) is seen to be
essentially identical to the $1s$  radial density  of a muon (energy
0) in the field of a nucleus with $Z=840, R=14.6$~fm (black curve).
Lower panel: The functions $E-mc^2 - V(r)$  for the $E=-mc^2$ case
(red curve) and $E=0$ case (black curve) are essentially identical
in the region where the densities are nonvanishing, explaining the
fact that the densities agree in the two cases.
 \label{fig2}
}
\end{figure*}

\begin{figure*}
\centerline{\includegraphics[scale=0.8]{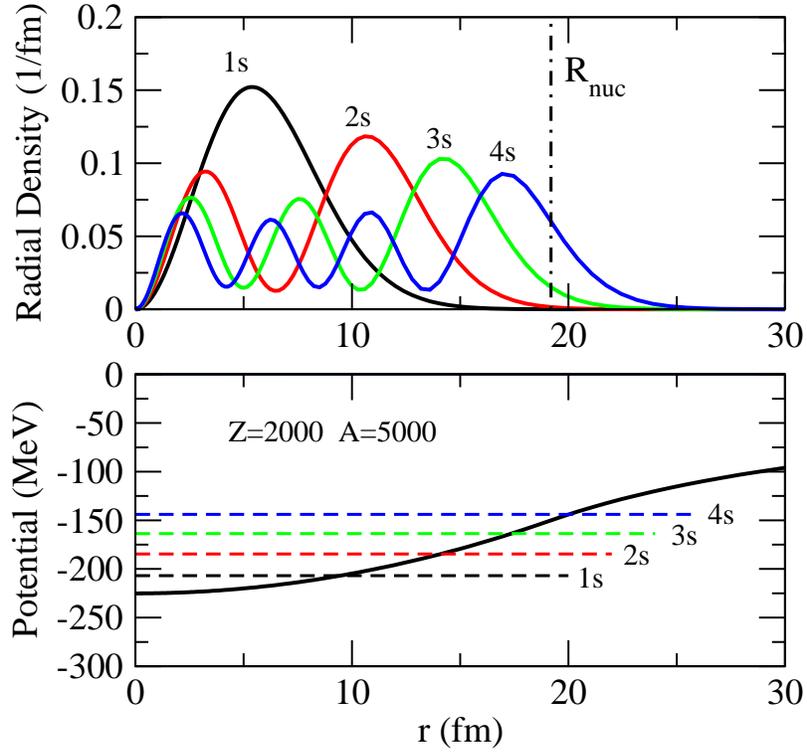}}
\vspace{1.5pc}

\caption{Upper panel: Radial densities for the four lowest $ns$
states of a muon in the field of a nucleus with charge $Z=2000$ and
radius $ R=19.2$~fm. Lower panel: Energies of the lowest four $ns$
states are  shown in comparison with the potential energy $V(r)$.
 \label{fig3}
}
\end{figure*}

 In the non-relativistic case, the radius of $1s$ state is inversely
proportional to the nuclear charge $Z$. Surprisingly, in the
relativistic case $E<0$,  the muon wave function is practically independent
 of $Z$ (see Fig.~\ref{fig2}, upper panel). The explanation is as
follows:  In the main area,
 the electrostatic potential $V(r)$ is approximately constant.
 A constant potential $V$ shifts the energy (by $V$), however, it does not
 change the wave function since the Dirac (and Schr\"{o}dinger) equations
contain the combination $E-V$  which  does not change.
 To illustrate this explanation,
 we plot $E-mc^2-V(r)$ for $Z=840$
 ($E\approx 0$) and for $Z=2000$ ($E\approx -mc^2$) in the lower panel of Fig.~\ref{fig2}
The corresponding wave functions practically coincide (see the upper
panel). In Fig.~\ref{fig3},  we present the radial densities of
$1s,\, 2s,\, 3s$ and $4s$ states for critical charge $Z_c\approx
2000$. All  of these states are localized inside the nucleus.

   Note that the ``critical charges'' for muons are so large that there is
hardly any  chance of actual stabilization of a ``normal'' nuclear matter
by muons. Therefore, in the muon case this problem is only an exercise
 in the relativistic quantum mechanics (critical charges in the oscillator-like
 potential of the finite nucleus). However, this solution may give us some
hints as to what one should expect in  more interesting cases of pions and kaons.
We do not consider these cases in the present note since they require
more sophisticated calculations including the strong interaction.

The authors are grateful to K. T. Cheng for help in the numerical
calculations. The work  of V.V.F. is supported by the Australian
Research Council. The work of W.R.J. is supported in part by NSF
grant No. PHY-04-56828.

\end{document}